# Correlated Insulating and Superconducting States in Twisted Bilayer Graphene Below the Magic Angle

Emilio Codecido[1], Qiyue Wang[2], Ryan Koester[1], Shi Che[1], Haidong Tian[1], Rui Lv[1], Son Tran[1], Kenji Watanabe[3], Takashi Taniguchi[3], Fan Zhang[2*], Marc Bockrath[1*], Chun Ning Lau[1*]

[1]Department of Physics, The Ohio State University, Columbus, OH 43210, USA

[2]Department of Physics, The University of Texas at Dallas, Richardson, TX 75080, USA

[3]National Institute for Materials Science, 1-1 Namiki Tsukuba Ibaraki 305-0044 Japan.

Email: zhang@utdallas.edu, bockrath.31@osu.edu, lau.232@osu.edu

ABSTRACT

The emergence of flat bands and correlated behaviors in "magic angle" twisted bilayer graphene (tBLG) has sparked tremendous interest, though many aspects of the system are under intense debate. Here we report observation of both superconductivity and the Mott-like insulating state in a tBLG device with a twist angle of ~0.93º, which is smaller than the magic angle by 15%. At an electron concentration of ±5 electrons/moiré unit cell, we observe a narrow resistance peak with an activation energy gap ~0.1 meV, indicating the existence of an additional correlated insulating state. This is consistent with theory predicting the presence of a high-energy band with an energetically flat dispersion. At a doping of ±12 electrons/moiré unit cell we observe a resistance peak due to the presence of Dirac points in the spectrum. Our results reveal that the "magic" range of tBLG is in fact larger than what is previously expected, and provide a wealth of new information to help decipher the strongly correlated phenomena observed in tBLG.

Twistronics[1-7], the use of the relative twist angle between adjacent van der Waals layers to produce a moiré superlattice and flat bands, has emerged as a new and uniquely suitable approach to dramatically alter and tailor the properties of devices based on two-dimensional (2D) materials. The dramatic effect of twistronics is exemplified by the recent groundbreaking works that demonstrated the emergence of extremely flat bands when two monolayer graphene layers are stacked at a "magic" twist angle, $\theta$ =1.1±0.1°[8, 9]. In these twisted bilayer graphene (tBLG) devices at the magic angle (MA), an insulating phase is observed at half filling of the superlattice's first miniband, identified to be a Mott-like insulator, and superconductivity at slightly higher and lower doping. The phase diagram is reminiscent of high-temperature superconductors[10], for the high ratio between the superconducting transition temperature $T_c$ to the Fermi temperature $T_F$, the relatively small Fermi surface, and the vicinity of the superconducting phase to an insulating state with an apparent magnetic ordering. These works have immediately sparked tremendous interest, and ignited intense theoretical debate on nearly every aspect of this system, including low energy band structure, band topology, irreducible symmetries, nature of the correlated insulating state, pairing mechanism and symmetry of the superconducting phase, the exact phase diagram, and additional magic angles[11-33]. In contrast, apart from the initial reports, experimental studies are scarce and just starting to emerge[33-37].

Here we report transport measurements of a MA-tBLG device that exhibits both correlated insulating and superconducting states. Surprisingly, the twist angle is measured to be 0.93°±0.01°, which is 15% smaller than the already established magic angles[7-9] and is the smallest reported to date that exhibits superconductivity. A correlated insulating state is observed at half filling $n_m$=±2, where $n_m$ is the number of charges per moiré unit cell, and superconductivity at $n_m$≈2.7. Extending measurements to carrier densities $|n_m|$>4, we observe previously unreported, narrow resistance peaks at $n_m$=±5, each of which displays an activation gap of ~0.1 meV. This behavior is consistent with the emergence of additional Mott-like correlated insulating states when the next electron or hole band beyond the low-energy moiré Dirac bands is quarterly filled.  Theoretical calculations indicate that these two high-energy bands have energetically flat regions in the moiré Brillouin zone with substantially large densities of states (DOS). Prominent resistance peaks also appear at $n_m$=±12, which can be accounted for by the existence of a pair of Dirac points in the high-energy spectrum. Our results indicate that new correlated states can emerge in tBLG below the primary magic angle and beyond the first miniband, provided the bands are sufficiently flat.

Devices are fabricated using the "tear and stack" approach, encapsulated between hexagonal BN layers[8, 9], patterned into a Hall bar geometry with multiple leads, and coupled to Cr/Au edge contacts[38]. The entire device is fabricated on top of a graphite layer that serves as the back gate (e.g. see ref. [39]). Figure 1(A-B) shows a schematic diagram of the layer stack and the moiré superlattice, respectively. The devices are then measured in pumped He[4] and He[3] cryostats using standard dc and ac lock-in techniques.

Figure 1C displays the device's longitudinal resistance $R_{xx}$ vs. an extended gate voltage $V_g$ range and magnetic field $B$ at a temperature $T$=1.7K, while the inset to Fig. 1D shows an optical image of the device. The main resistance peak at $V_g$=0 corresponds to the charge neutrality point with $n_m$=0. From the Landau fan emanating from $V_g$=$B$=0, we find the back gate capacitance to be $C_b$ = 374 nF/cm$^2$. Notably, two additional prominent peaks appear at $V_g$=±0.85 V, each accompanied by a set of Landau fans. We therefore take the features at $V_g$=±0.85 V as the satellite peaks that occur when the low-energy moiré bands are filled at densities $n_m$=±4 [8, 9]. Covering an unprecedentedly large range of carrier density up to $n_m=\pm14$, the data exhibit Landau fans that converge to $n_m=4m$, where $m$ is an integer. Given the spin and valley degeneracies in graphene, this is consistent with the Wannier theory[40] that predicts generally that spectral gaps in the Hofstadter butterfly for spinless electrons in a single band follow the Diophantine equation $n_m=t\phi/\phi_0+s$, where $s$ and $t$ are integers, $\phi$ is the flux per moiré unit cell, and $\phi_0$ is the flux quantum. Using $4n_0 \approx \frac{8\theta^2}{\sqrt{3}a^2}$, where $a$=0.246 nm is the lattice constant of graphene and $n_0$ is the density corresponding to $n_m$=1, we estimate that $\theta$=0.93º, which corresponds to a moiré unit cell area $A$ of 200 nm$^2$ and moiré lattice constant of 15.2 nm. We note that this is the smallest twist angle value reported to date for tBLG devices exhibiting superconductivity.

A close examination of the Landau fan in Fig. 1C reveals a number of salient features. We first focus on the low-to-moderate density regime, where $|n_m|<4$. At $V_g$=0.43V or $n_m$=+2, a resistance peak appears, from which an accompanying set of Landau levels emanates, with a degeneracy of two. This peak at half filling and the two-fold degeneracy of Landau levels are consistent with the prior observation of a Mott-like correlated insulating state[8]. They indicate the breaking of spin-valley SU(4) symmetry and the formation of a new quasiparticle Fermi surface, although the details require more delicate examination. Figure 1D displays $R_{xx}(V_g)$ at $B$=0 and $T$ ranging from 5.2 K down to 0.28 K, where the resistance peaks at $n_m$=0, ±4 and +2 are visible. Interestingly, at $T$=280 mK, for 0.51<$V_g$<0.65, or equivalently, 2.4< $n_m$<3.1, $R_{xx}$ is zero within our measurement error, indicating the emergence of superconductivity[9].

To determine the critical temperature $T_c$ of the superconducting phase, we measure $R_{xx}(T)$ at $V_g$=0.53V or $n_m$≈2.5 (Fig. 2A). As $T$ decreases, $R_{xx}$ drops to zero, undergoing two successive steep descents at $T$~1.5 K and $T$~0.3 K. Such a two-step transition have been observed in other magic angle tBLG device[34], and may be related to non-Planckian dissipation of the strange metal state. Alternatively, it may arise from spatial or structural inhomogeneity of our device, or from the presence of domains that host competing superconducting states of different pairing symmetries and critical temperatures[41].

To further investigate the superconducting phase, we measure the device's four-terminal voltage-current (V-I) characteristics at different carrier densities. Fig. 2B displays resistance, which is obtained by numerically differentiating the V-I curves, as a function of $V_g$ and bias current $I$. V-I characteristics at two representative densities, $V_g$=0.50 V and $V$=0.58 V, are shown in Fig. 2C. Supercurrent is observed for an extended range of density, with critical current $I_c$ ranging from ~ 1 nA to 15 nA; at $V_g$~0.58 V, the maximum value of $I_c$ is observed. We therefore take $V_g$~0.58 V (or $n_m$~2.7) to be the optimal doping. Upon application of a parallel magnetic field $H$, the supercurrent is suppressed, with a critical field of $H_{||c}$ ~ 0.5T (Fig. 2D), consistent with previous work[9].

To gain insight into this behavior, we calculate the moiré band structure for the 0.93º tBLG using the Bistritzer-MacDonald model[7] with refined parameter values that take into account lattice relaxation: $t_{AA}$ = 97.5 meV, $t_{AB}$ = 79.7 meV, and $v_F$ = 7.98×10$^5$ m/s [42]. Here $t_{AA}$ and $t_{AB}$ are the electron interlayer tunneling amplitudes between the same and different sublattices, respectively, and $v_F$ is the Fermi velocity. In sharp contrast to those cases near and above the magic angle, our calculation of the 0.93º t-BLG shows that the low-energy moiré Dirac bands ($|n_m|<4$) are not energetically isolated from the high-energy bands. While the single-particle superlattice gaps vanish at the complete filling of the low-energy moiré Dirac bands, new Dirac points exist at $\Gamma_s$, as shown in Fig. 3A. This explains why our $R_{xx}$ peaks at $n_m=\pm 4$ are narrower compared to a prior report[8] and comparable to the one at the charge neutrality. The fact that the $R_{xx}$ peak is relatively stronger at $n_m=4$, together with the fact that superconductivity emerges only at the electron side, suggests that the electron-hole asymmetry is substantially enhanced by the electron-electron interactions. The emergence of both the Mott-like correlated insulating state at half-filling and the superconductivity at slightly higher doping indicates that, though the twist angle of 0.93º is ~15% smaller than the magic angle[7-9], the device indeed hosts strongly correlated physics. This unexpected yet desirable behavior can also be understood by the calculated moiré bands and DOS of the 0.93º tBLG in Fig. 3. The low-energy moiré Dirac bands ($|n_m|<4$) are bounded by the aforementioned multiple high-energy Dirac points near -4.68 meV and 5.96 meV. The narrow bandwidth (~11 meV), comparable to that of the magic-angle tBLG [8, 9] and much smaller than the Coulomb interaction strength, produces a sharp DOS peak for $|n_m|<4$.

We now turn to the behavior at large density ($|n_m|>4$). At the lowest temperature, additional narrow resistance peaks are observed at $n_m=\pm 5$ (Fig. 1c). Figure 4A shows a zoomed-in plot of the data for one of the peaks. The peaks are almost indiscernible as $T$ is increased above ~5 K. Plotting the resistivity on an Arrhenius plot as shown in Fig. 4B yields an energy gap ~1.3 K. These resistance peaks at $n_s=\pm 5$ have not been previously reported. Conceivably, it may arise from the presence of another domain with a slightly larger angle. Indeed, domains are likely to be present[33]. However, the sharpness of the peaks and their small energy scale are very different from the behavior expected for a superlattice gap[8, 9], which are much broader and show very little low-temperature dependence. Thus, the peaks at $n_s=\pm 5$ are unlikely to originate from angular disorder. We thus tentatively attribute these features to the emergence of a new correlated insulating state when the lowest (highest) high-energy conduction (valence) band is quarterly filled with electrons (holes). Evidently, in Fig. 3, our calculation reveals that these two bands are nearly flat in a large region of the moiré Brillouin zone, and that the corresponding DOS peaks are substantially large. From a background-subtracted $R_{xx}(V_g, B)$ plot, several Landau levels can be observed emanating from the two peaks and disperse toward the larger density sides, with degeneracy estimated to be 10±2, where the relatively large error bar of ±2 arises from the limited range in magnetic field at which the features clearly extrapolate to $n_m=\pm 5$ and $B=0$. Hence, within error bars, the degeneracy measured is larger than the band degeneracy of 4. This unusual feature suggests that the fermionic quasiparticles not only respect the spin-valley SU(4) symmetry but might even enjoy an emergent new symmetry as well. More delicate future studies are required to examine the symmetries of these exotic insulating states and to determine whether or not they are quantum spin liquids.

Another unusual feature of the device behavior at high density is the presence of resistance peaks at $n_m=\pm 12$, where Landau fans with degeneracy of 4 emanate on both sides of each density. Additionally, their maximum resistivities yield minimum conductivities ~ $20e^2/h$. These features are reminiscent of the presence of Dirac points at the charge neutrality. In fact, our moiré band structure calculation of the 0.93º tBLG does reveal that a pair of Dirac points does appear at $K_s$ and $K'_s$ near $n_m=\pm 12$ (Fig. 3). Consistently, the DOS calculation also shows local minima at the corresponding energies (Fig. 3).

In sum, in a twisted bilayer device at a small twist angle of 0.93º, superconductivity is observed near a Mott-like insulating state, with a critical temperature of 0.3-0.5K. A gap is observed at a filling of ±5 electrons per moiré unit cell. Theoretical calculations predict no band gap at this filling, indicating correlated insulating behavior for the quarter filled high-energy band. Dirac points at a filling of ±12 electrons per moiré unit cell lead to conductivity minima. Our work shows that electron correlations can dramatically influence the properties of moiré superlattices even at small angles and high densities. Future work will focus on the spin-valley ordering of the insulating phases and investigations at lower temperatures to search for new superconducting phases, as well as theoretical efforts to understand the origin of these behaviors.

**Acknowledgement:** The experiments are supported by DOE BES Division under grant no. ER 46940-DE-SC0010597. The theoretical works (QW and FZ) are supported by Army Research Office under Grant Number W911NF-18-1-0416. FZ is grateful to Fengcheng Wu and Adrian Po for valuable discussions. Growth of hBN crystals was supported by the Elemental Strategy Initiative conducted by the MEXT, Japan and a Grant-in-Aid for Scientific Research on Innovative Areas "Science of Atomic Layers" from JSPS. We thank the groups of J. Hone and C. Dean for experimental advice on device fabrication.

**Fig. 1. Device geometry and magneto-transport data.** (**A**) Schematic diagram of device geometry. (**B**) Schematic diagram of moiré superlattice formed by the twisted graphene layers. (**C**) $R_{xx}$ vs. magnetic field $B$ and gate voltage $V_g$ showing a Landau fan pattern. The top axis labels $n_m$, the number of charges per superlattice cell. (d). $R_{xx}(V_g)$ at different temperatures. Inset: optical image of a tBLG device.

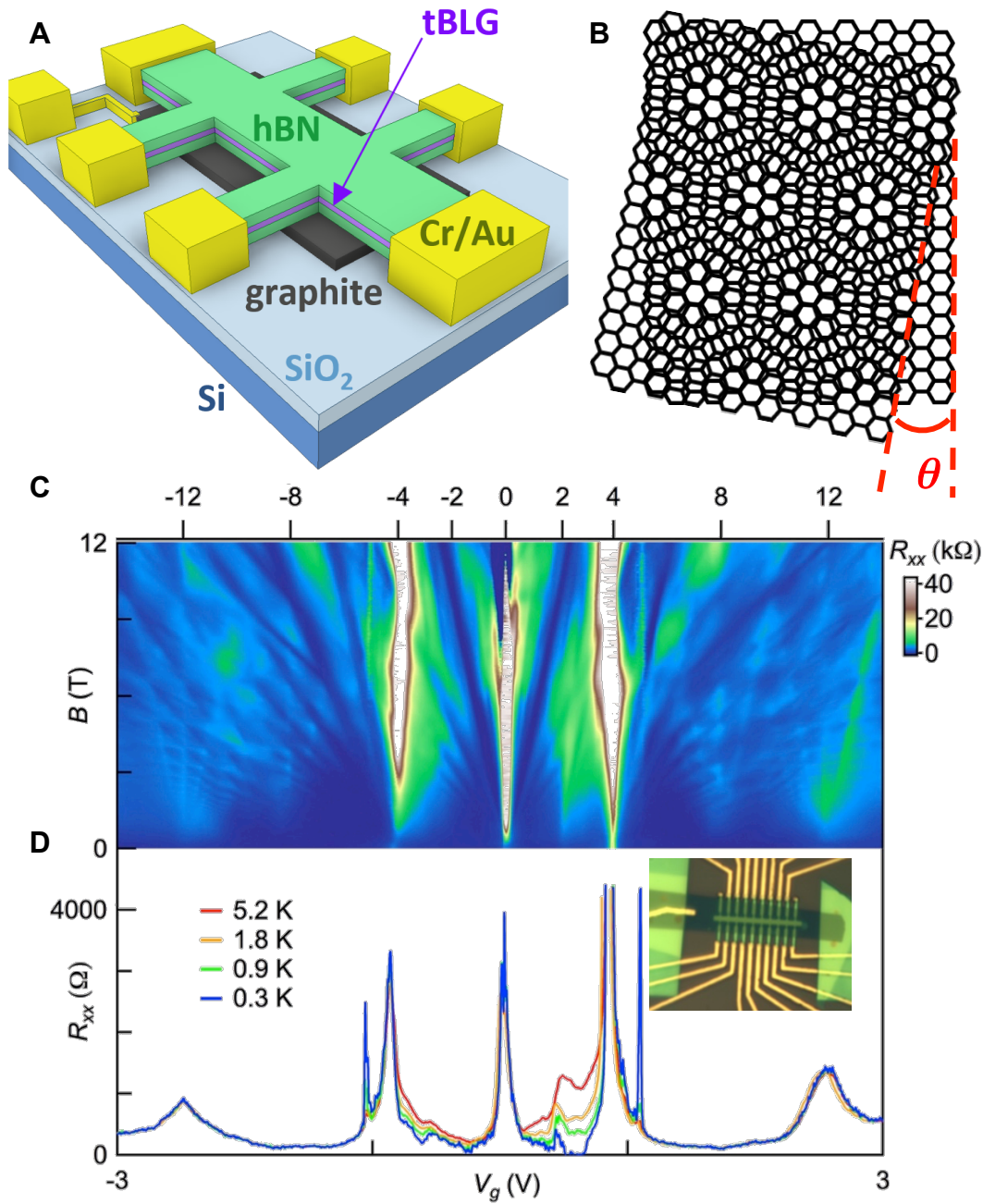

**Fig. 2. Data from the superconducting state.** (**A**) $R_{xx}$ vs. temperature when the density is tuned to the superconducting phase ($V_g$~0.53V, or $n_m$~2.5). (**B**) Differential resistance $dV/dI$ vs bias current and gate in the superconducting phase at base temperature (280 mK). Color scale is in units of kΩ. (**C**) Voltage-current characteristics at $T$=280mK and $V_g$=0.50 (blue) and 0.58V (red), respectively. (**D**) V-I curves at different parallel magnetic fields.

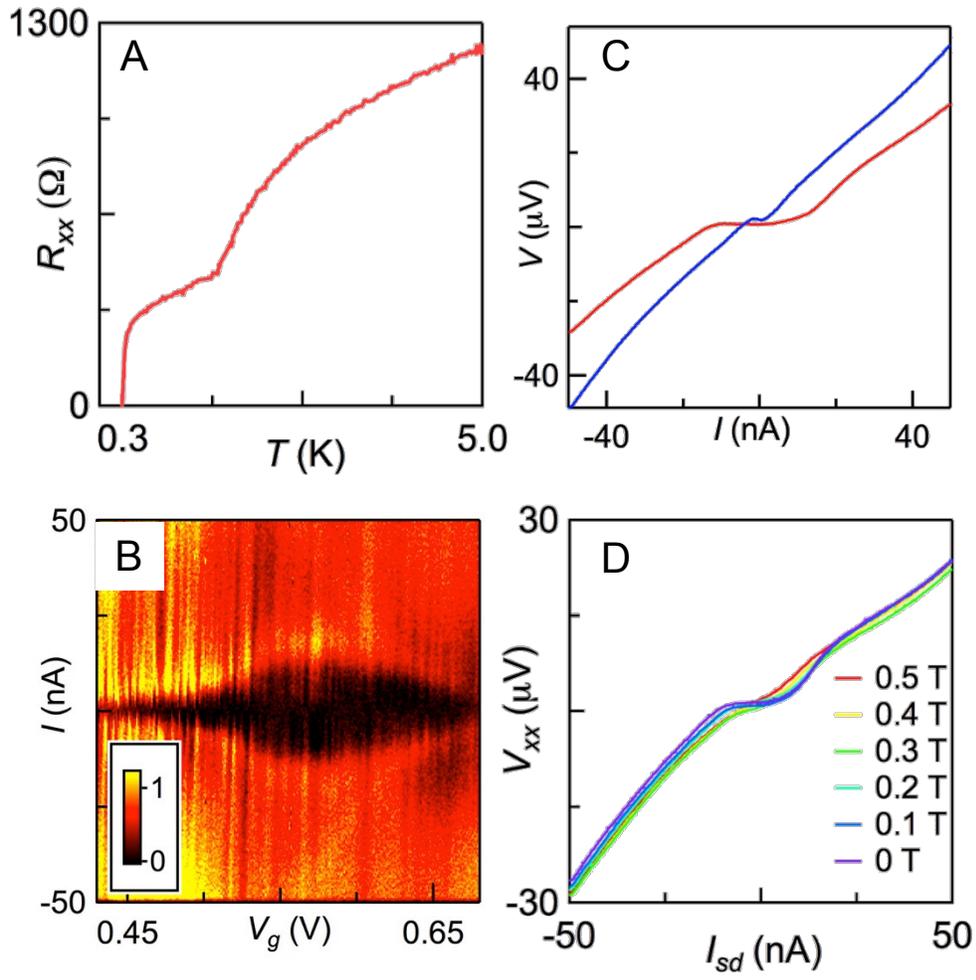

**Fig. 3. Theoretical calculations of** (**A**) band structure at valley *K* and (**B**) total density of states for the tBLG with $\theta$=0.93°. In obtaining the density of states from the band structure, 1 meV was used for the energy interval, and the spin-valley degeneracy was considered.

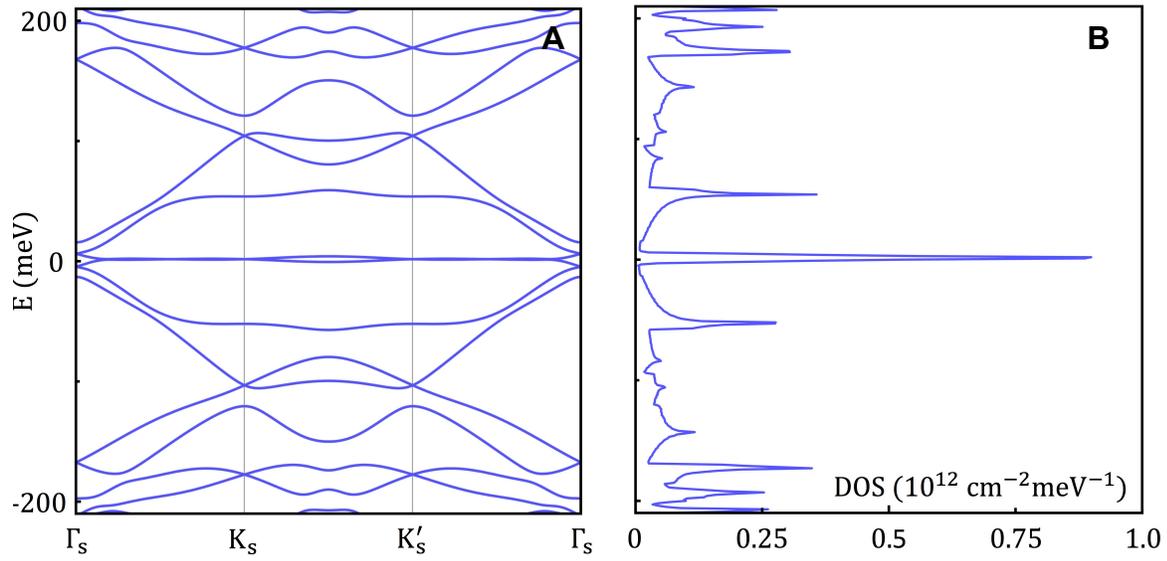

**Fig. 4. Behavior of resistance peak near density** $n_m$=**5.** (**A**) Temperature dependence of the resistance peak. (**B**) Arrhenius plot of resistance showing a gap ~0.1 meV.

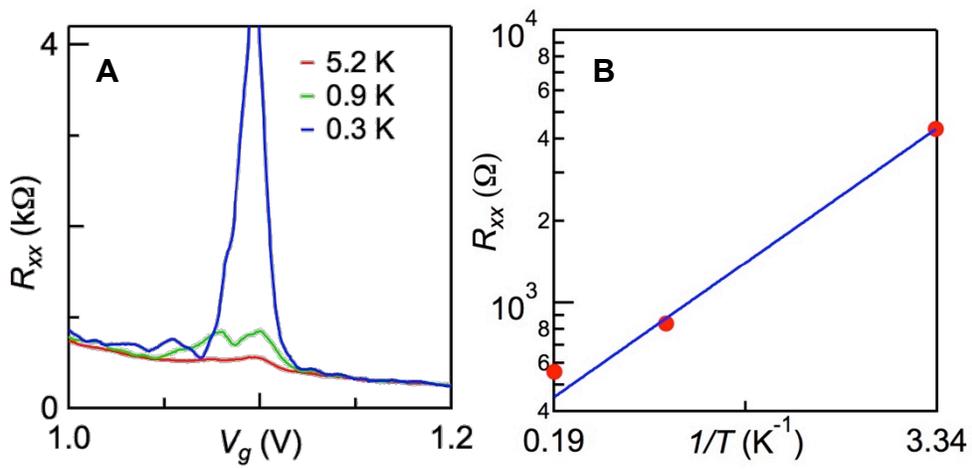